\documentclass{aa} 
\let\linenumbers\relax
\usepackage{natbib}
\bibpunct{(}{)}{;}{a}{}{,}
\usepackage[varg]{txfonts}
\usepackage{graphicx}
\usepackage[colorlinks=true,linkcolor=black,anchorcolor=black,citecolor=blue,filecolor=black,menucolor=black,runcolor=black,urlcolor=black]{hyperref}

\begin{document}
\title{Magnetic evolution of complex solar active regions}
  
  \author{Shabnam Nikbakhsh\inst{1,2}
  \and Eija I. Tanskanen\inst{1,2} 
  \and Thomas Hackman \inst{3}} 

\institute{Sodankylä Geophysical Observatory, University of Oulu, Oulu, Finland \\ email:shabnam.nikbakhsh@oulu.fi
 \and Department of Electronics and Nanoengineering, Aalto University, P.O. Box 15500, FI-00076, Aalto, Finland
  \and Department of Physics,  P.O. Box 64, FI-00014, University of Helsinki, Finland} 


\abstract {}{This study investigates the magnetic evolution of solar active regions (ARs), with a particular focus on understanding how the magnetic morphology of simple and complex ARs changes throughout their lifetime.}{To analyse the magnetic evolution of ARs, we developed a Magnetic Evolution Method (MEM) that segments each region’s lifetime into three phases: growth, main, and recovery. The method was applied to ARs observed between January 1996 and December 2020.}{We found that complex active regions (CARs) have a mean lifetime of approximately 24 days, which is 8 days longer than that of simple active regions (SARs). Most CARs ($94\%$) first appear with a simple magnetic structure and remain in this configuration for about 3 days (growth phase), before transitioning into complex structures for around 5 days (main phase), after which they typically revert to a simple state (recovery phase). The average lifetimes of SARs and CARs show no significant difference between solar cycles 23 and 24, suggesting that active region lifetimes are independent of the solar cycle.}{By tracking the full magnetic evolution of ARs, our study reveals that CARs typically become magnetically complex around three days after emergence and remain in that state for a limited but critical period. This temporal structure, uncovered using a novel method that follows ARs throughout their full development, provides important context for identifying the magnetic conditions associated with increased eruptive potential. The results offer a foundation for improving the forecasting of solar flares and magnetic clouds, and suggest that the magnetic evolution of ARs is largely independent of the solar cycle.}

\keywords{Sun: magnetic fields – Sun: Active Regions - Sun: solar flare}
\maketitle 

\section{Introduction}
\label{sec:intro}
The solar photosphere is the location where active regions (ARs) emerge with different magnetic field strengths and magnetic complexities \citep{Howard1989, Howard1991}.  A substantial number of studies have attempted to classify ARs based on their magnetic structures or complexities \citep{Cortie1901, Hale1919, Waldmeier1938, McIntosh1990}. Among all the magnetic classifications, the Mount Wilson Scheme has been broadly applied in studying the magnetic structure of ARs \citep{Tang1984, Ireland2008, Stenning2013}. The modified Mount Wilson Scheme classifies ARs into the five major groups: $\alpha$, $\beta$, $\beta\gamma$, $\gamma$ and $\delta$ \citep{Van2015}. When an active region emerges into the photosphere, it might present one or a combination of two magnetic classes during its lifetime. ARs can also produce solar flares and coronal mass ejections (CMEs). The production rate of flares and CMEs in ARs has been shown to be higher in more magnetically complex regions \citep{Chen2011, Guo2014}. Although not all flares and CMEs originate within ARs, studying the magnetic class of ARs could significantly improve space weather forecasting \citep{Subramanian2001}.

\citet{Jaeggli2016} studied the Mount Wilson classification of 5468 ARs from January 1992 to December 2015. This period includes the end of solar cycle (SC) 22, the entire period of SC 23, and  SC 24 up until 2015. They utilized the NOAA Active Region (NAR) database, provided by National Oceanic and Atmospheric Administration (NOAA) in order to study the cyclic variation in the magnetic complexity of ARs. The magnetic complexity of each AR was selected when the region achieved its maximum area. Hence, the variations in the magnetic complexity of ARs through the lifetime of the regions were not considered. They reported that the majority of ARs in their sample had simple magnetic structures, which are $\alpha$ and $\beta$ classes, while only $16.24 \%$ of ARs emerged with complex structures. 

In a more recent study, \citet{Nikbakhsh2019} also investigated the solar cycle variability of 4797 ARs for the period of December 1992 to January 2018, covering the whole SC 23 and almost the entire SC 24.  They studied the Solar Active Region Summary (SRS) list provided by NOAA; however, \citet{Nikbakhsh2019} applied a different method than \citet{Jaeggli2016} in order to examine the magnetic complexity of ARs. \citet{Nikbakhsh2019} selected the magnetic complexity of each AR per day and calculated the daily abundance of each complexity class. By applying this method, they studied the cyclic variation of ARs and compared their results to the daily sunspot number, which is also calculated for each day \citep{Clette2016}. \citet{Nikbakhsh2019} reported that the majority of the regions in their sample had simple magnetic structures, while about $12 \%$ had complex configurations. 

The magnetic complexity of ARs has been widely studied since the discovery of magnetic field presence in sunspots \citep{Hale1908}. However, to the best of our knowledge, none of the previous studies have statistically investigated the magnetic evolution of ARs during their lifetime. In this study, we introduce a new approach to track daily changes in magnetic structure, enabling a detailed statistical analysis of how ARs transition between simple and complex morphologies. This method provides new insights into the development of complex regions and their potential role in solar activity.

\section{Data and method}
\label{sec:DATA}

We studied the magnetic complexity of ARs present in the SRS list by applying our new Magnetic Evolution method (MEM) for the period of January 1996 -- December 2020, which covers the end of SC 22 , and the entirety of SC 23 and 24. The SRS list is provided by NOAA and the Solar Optical Observing Network (SOON), and it reports the daily observations of ARs present on the visible solar disk for the preceding day \citep{Andrus2013}. The list includes the daily data of ARs, such as the Mount Wilson and Zurich magnetic classifications, total area, heliographic longitude and latitude, and number of sunspots for each individual region \footnote{{https://www.helio-vo.eu/}}. The Mount Wilson magnetic classification reported in the SRS list comprises $\alpha$, $\beta$, $\gamma$, $\gamma\delta$, $\beta\delta$, $\beta\gamma$, and $\beta\gamma\delta$ classes.

In order to investigate the complexity data of ARs, we divided them into two separate groups utilizing our MEM: simple active Regions (SARs) and complex active regions (CARs); If an AR appeared on the visible solar disk as a $\gamma$, $\gamma\delta$,$\beta\delta$, $\beta\gamma$, or $\beta\gamma\delta$ region during its lifetime, even for one day, we labelled that region as a CAR. On the other hand, if an AR appeared only as a $\alpha$ or $\beta$ region during its entire lifetime, we marked it as a SAR. Furthermore, we compared our results to the Boulder/NOAA sunspot number, which we term NSN. Note that NSN is computed in a similar manner as the international sunspot number (ISSN), and these sunspot numbers closely follow each other \citep{Hathaway2015, Clette2016}. Our method in studying the magnetic complexity of ARs is different from \citet{Jaeggli2016} in that they only considered the magnetic complexity of ARs when regions reached their maximum area, while we studied the Magnetic Evolution of each individual region during its entire lifetime. Furthermore, our method is also different from \citet{Nikbakhsh2019} in the way that they determined the magnetic complexity of ARs for each day. Thus, if an AR is classified as simple for a day, it could be classified as complex for another day. Hence, their approach is more favourable towards long-lived ARs rather than short-lived ones. 

We also calculated the lifetime of each individual AR in our dataset. The SRS list provides both the first and last observation dates of each region on the visible solar disk, allowing us to determine how many days each AR was observed and what magnetic complexity it possessed on each day. Notably, an AR may persist for multiple solar rotation periods. Our approach differs from that of \citet{Schrijver2000}, who defined the lifetime of an AR only while it remained a bipolar system. In contrast, we consider the entire period during which the region was visible. Our method is more closely aligned with that of \citet{Nagovitsyn2019}, who defined sunspot group lifetimes based on the number of days of observation.	

By examining each AR throughout its entire lifetime, our method enables a detailed analysis of how magnetic structures evolve over time, distinguishing between SARs and CARs. This approach allows us to quantify the timing and duration of magnetic complexity, offering a new perspective on the dynamic evolution of ARs. Although our study does not directly assess eruptive events, the resulting insights into magnetic behaviour may contribute to future efforts in space weather forecasting.

\section{RESULTS}
\label{sec:RESULTS} 
We applied MEM to the SRS dataset to study the magnetic evolution of ARs over the period January 1996 to December 2020. Based on their magnetic structure during their lifetime, we identified a total of 4841 regions, of which 3943 were classified as SARs and 898 as CARs. Table  \ref{number} presents the total counts and relative abundance of both groups.

\begin{table}[ht]
\caption{Total number and abundance of SARs and CARs in the SRS list from January 1996 to December 2020.}
\label{number}
\centering
\begin{tabular}{ c  c  c  }
\hline\hline
  Complexity  & Number & Abundance  \\
  &  &  [$\%$]  \\
\hline      
SARs  & 3943  & 81.5 \\
CARs  & 898 & 18.5	 \\
\hline 
&  &    \\
Total ARs  & 4841 &  \\    
\hline  
\end{tabular}
\end{table}

 We further analysed these results to investigate the cyclic variation in the number of SARs and CARs. The number of SARs, CARs, and all ARs in SC 23 and 24 is compared in Table \ref{cycle}. The ratio of CARs to SARs  is  also presented in this table. As can be observed from the data in Table \ref{cycle}, the number of SARs decreased by $45.1 \%$ while the population of CARs dropped by only $26.0 \%$ from SC 23 to SC 24.  In addition, the total number of ARs decreased by $41.9 \%$ while the ratio of CARs to SARs vary $25.9 \%$ from SC23 to 24.

\begin{table}[ht]
\caption{Cyclic variation of SARs and CARs in SC 23 and 24.}
\label{cycle}
\centering
\begin{tabular}{ c  c  c  c } 
\hline\hline     
Complexity  & SC 23& SC 24 & Rate of change \\
&[number]  & [number] & [$\%$] \\
\hline      
SARs  & 2546  & 1397 &  45.1\\
CARs  & 516  &  382 &  26\\
Total ARs  & 3062 &   1779 &  41.9 \\
\hline 
&  &  &  \\
CARs/SARs ratio  & 0.20 &  0.27 & 25.9\\	
\hline 
\end{tabular}
\end{table}

\subsection{Lifetime of ARs}
We studied the mean lifetime of SARs and CARs in both SC 23 and 24. In order to do that, we calculated the lifetime of each individual region from the day it emerged into the visible disk until the last day it was observed. 
\begin{table}[h]
\caption{Mean lifetime of  SARs and CARs in SC 23 and 24.}
\label{Lifetime}
\centering
\begin{tabular}{ c c c c }
\hline\hline
&  SC 23 & SC 24 &  Both cycles\\
&[Day] & [Day] & [Day]\\
\hline      
SARs  & 15.9 &  15.1 &  15.6\\
CARs  & 24.8 &  22.3 & 23.8\\
\hline         
\end{tabular}
\end{table}

\begin{figure}[ht]
\resizebox{\hsize}{!}{\includegraphics{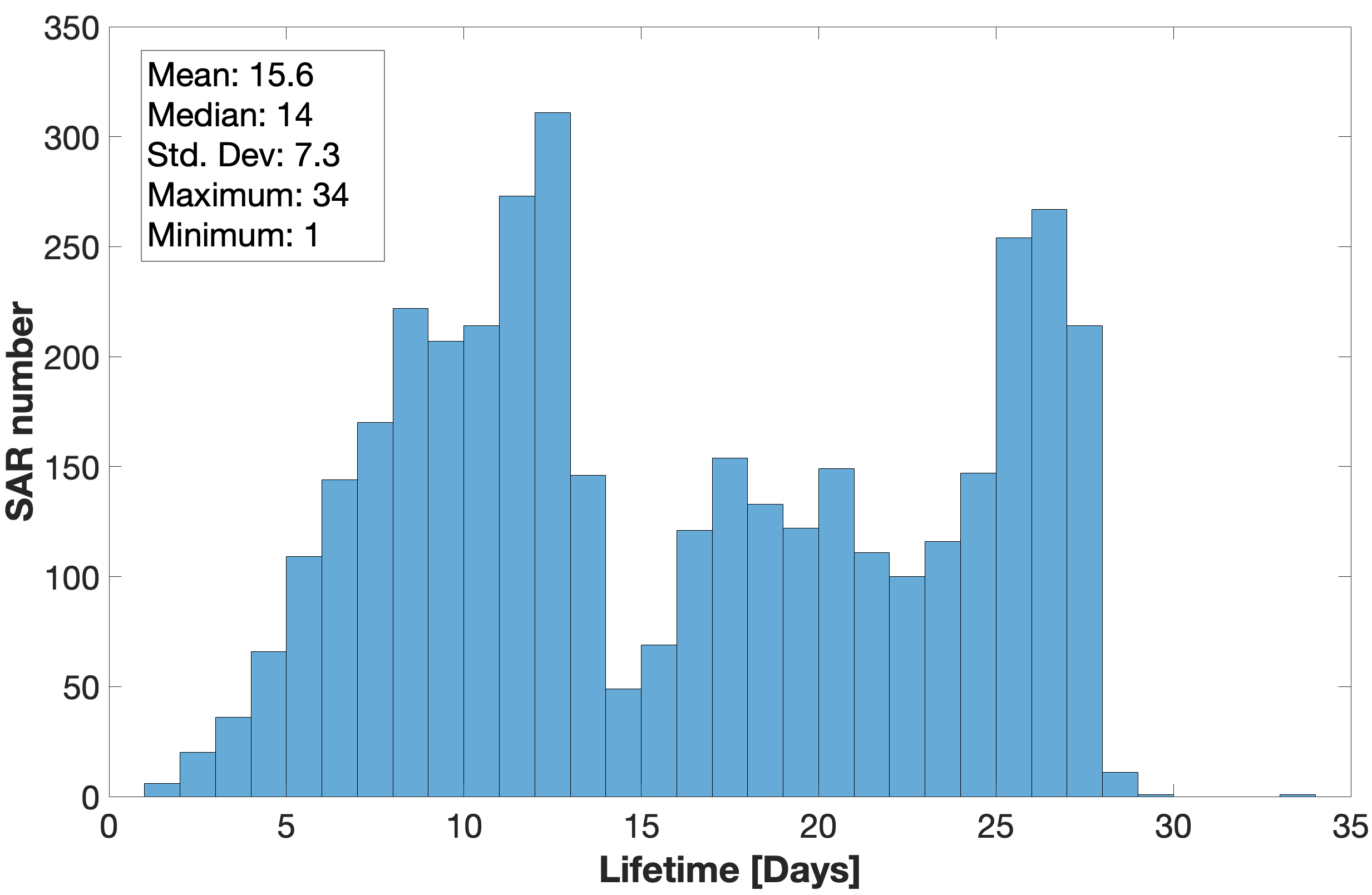}}
\\[1\baselineskip]
\resizebox{\hsize}{!}{\includegraphics{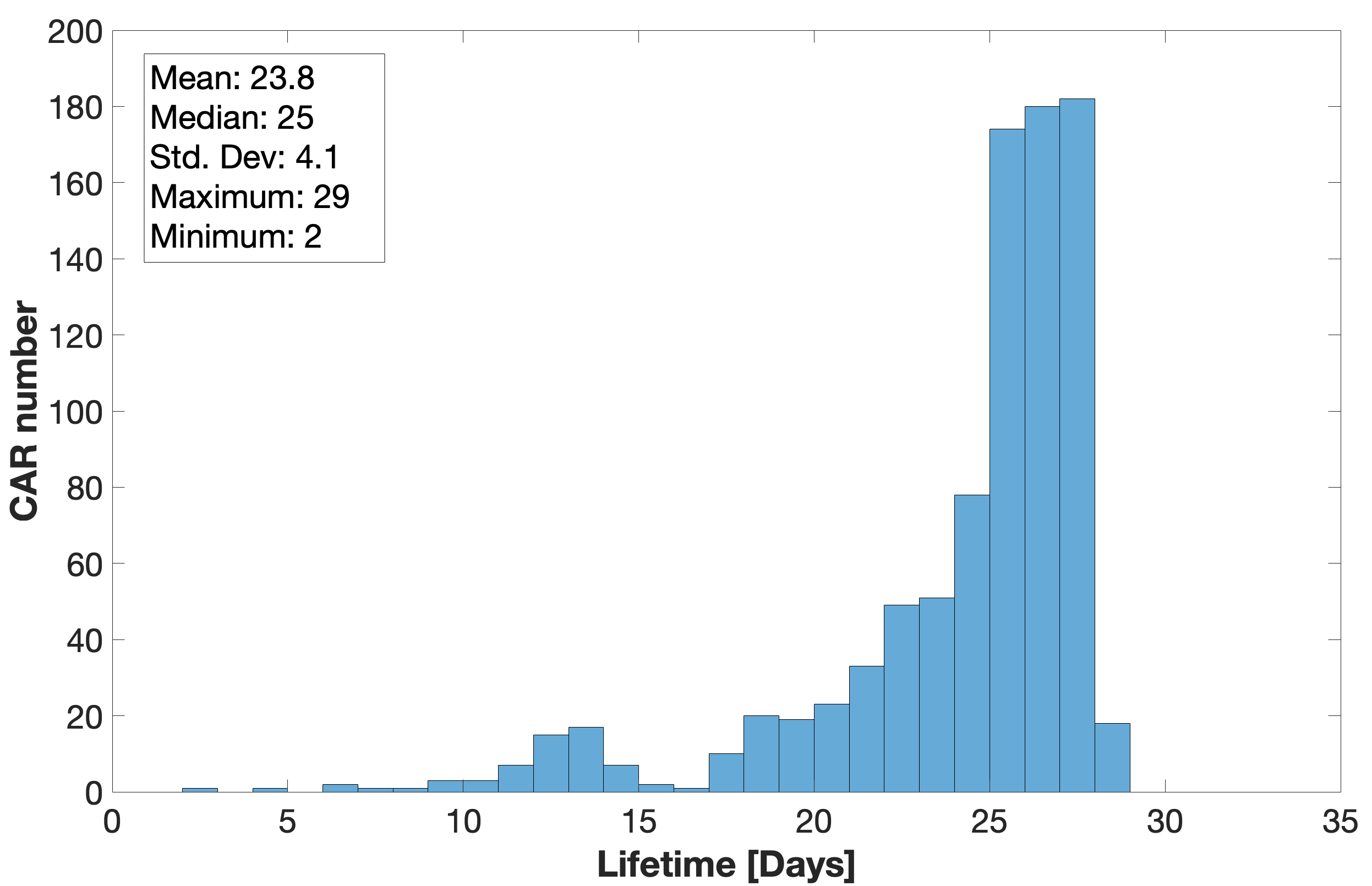}}
\caption{Lifetime distribution of SARs (upper panel) and CARs (lower panel) from January 1996 to December 2020.}
\label{lifetime}
\end{figure}

The observed mean lifetime of SARs and CARs in both SC 23 and 24 is presented in Table \ref{Lifetime}. In the table, it can be noted that CARs generally have a longer mean lifetime than SARs in both cycles (about 8 days longer).  In addition, the lifetime of SARs and CARs do not significantly vary from cycle 23 to cycle 24. 

For more detailed analysis on the lifetime of SARs and CARs, we calculated the lifetime distribution of each group separately for the entire period. The summary statistics and lifetime distributions of the SARs and CARs are illustrated in upper and lower panels in Fig \ref{lifetime}, respectively. As can be observed from the upper panel of the figure, the lifetime of SARs indicates a markedly bimodal distribution around 12 and 26 days and lifetime ranges from 1 to 34 days. Thereafter, we analysed the lifetime distribution of SARs for SC 23 and 24 separately in order to investigate the cycle dependency of the distribution shape. Similarly, we identified that the lifetime distribution of SARs was bimodal in both SC 23 and 24. 

As demonstrated in the lower panel of Fig. \ref{lifetime}, the lifetime of CARs has a left-skewed distribution with $87.8 \%$ of these regions having a lifetime longer than 20 days, although the range of the lifetime of CARs varies from 2 to 29 days. Moreover, we also found out that the distribution of CARs was left-skewed in both SC 23 and 24. 

\subsection{Magnetic evolution of CARs}
We investigated all 898 CARs present in our sample, and found out that 842 of them ($93.8 \%$) first emerged to the photosphere as a $\alpha$ or $\beta$ regions before they turned into complex structures, such as $\beta\gamma$ or $\beta\gamma\delta$. Hence, we further studied these 842 regions for the period in which they first emerge to the surface and had simple magnetic structures ($\alpha$ or $\beta$), we term this phase the Growth phase. Next, we considered the period in which these regions had complex structures (after the Growth phase and before decaying to simple magnetic structures), and we term this phase the Main phase. We also defined the Recovery phase in which CARs turn to be simple again after the Main phase. The results from this analysis are presented in Table \ref{evolution}.

\begin{table}[ht]
\caption{Magnetic evolution of CARs.}
\label{evolution}
\centering
\begin{tabular}{c c c c}
\hline\hline
 & Range     & Mean value & Standard  \\
 &  [Day]   &  [Day]   &  deviation\\
\hline      
 & & & \\  
Growth phase & 1 - 11 &  3.2 & 2.1\\
 & & & \\ 
\hline 
 & & &\\      
Main phase & 1 - 13 &  4.8 & 3.3\\
 & & & \\  
\hline 
 & & & \\     
Recovery phase & 0 - 25 &  15.9 & 3.6 \\ 
\hline            
\end{tabular}
\end{table}

It can be noted from the data in Table \ref{evolution} that the Growth phase of complex  regions varies from 1 to 11 days with the mean value of 3.2  days. This means on average CARs stay in Growth phase only for about 3 days. Further analysis  showed that $78.9\%$ of CARs had Growth phases shorter than 4 days, which considering the mean lifetime of CARs (about 24 days, Table \ref{Lifetime}) is a very short period. In addition, one can see from Table \ref{evolution} that the Main phase ranges from 1 to 13 days with the mean value of 4.8. Hence, on average CARs have complex structures for one-fifth of their lifetime. Furthermore, it is evident that the mean value of the Recovery phase for these regions, 15.9 days is almost the same as the mean lifetime of SARs (15.6 days, Table \ref{Lifetime}), although the recovery phase can vary between 0 to 25 days.

\subsection{Monthly number of SARs and CARs}
\begin{figure}[ht]
\resizebox{\hsize}{!}{\includegraphics{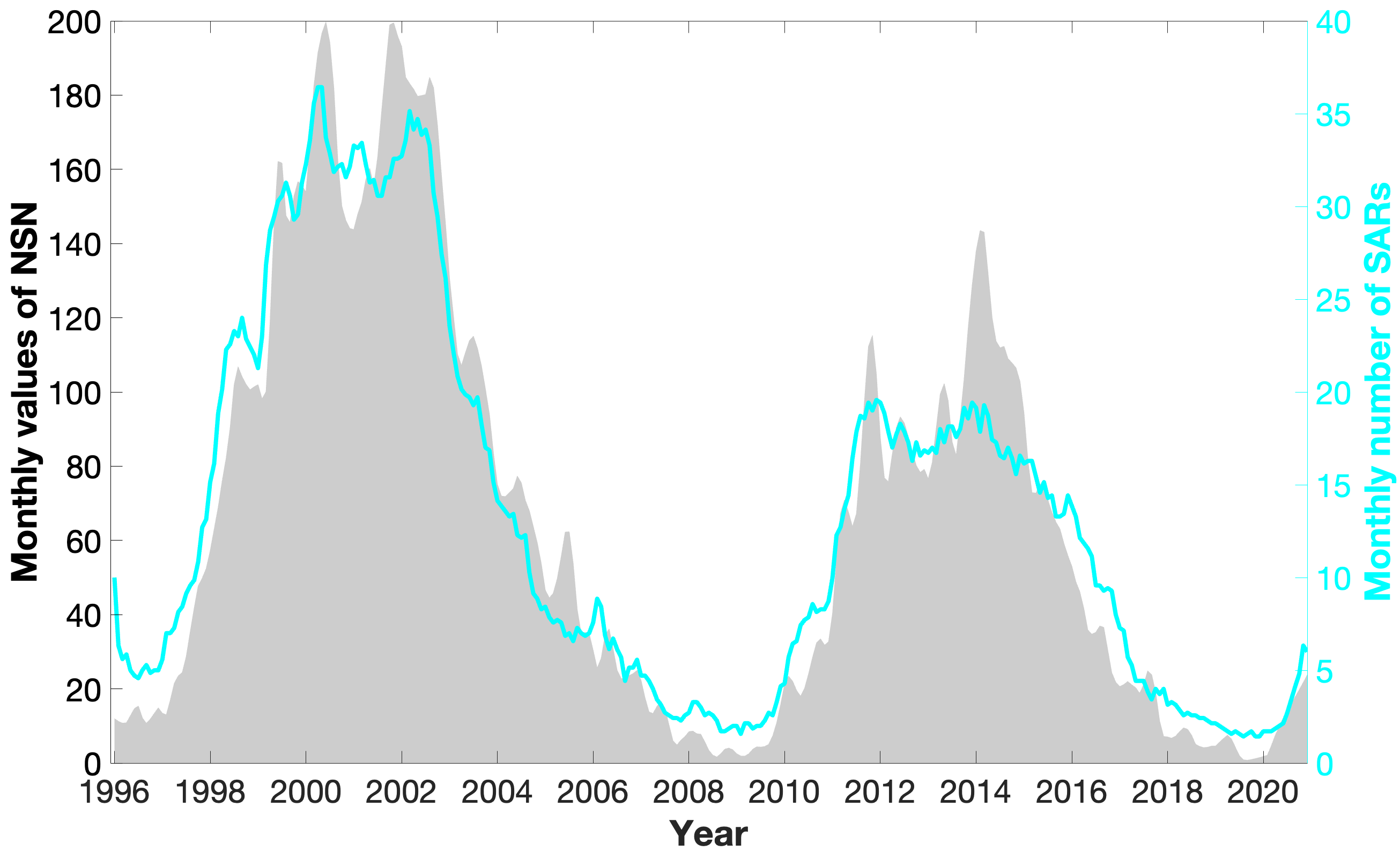}}
\\[1\baselineskip]
\resizebox{\hsize}{!}{\includegraphics{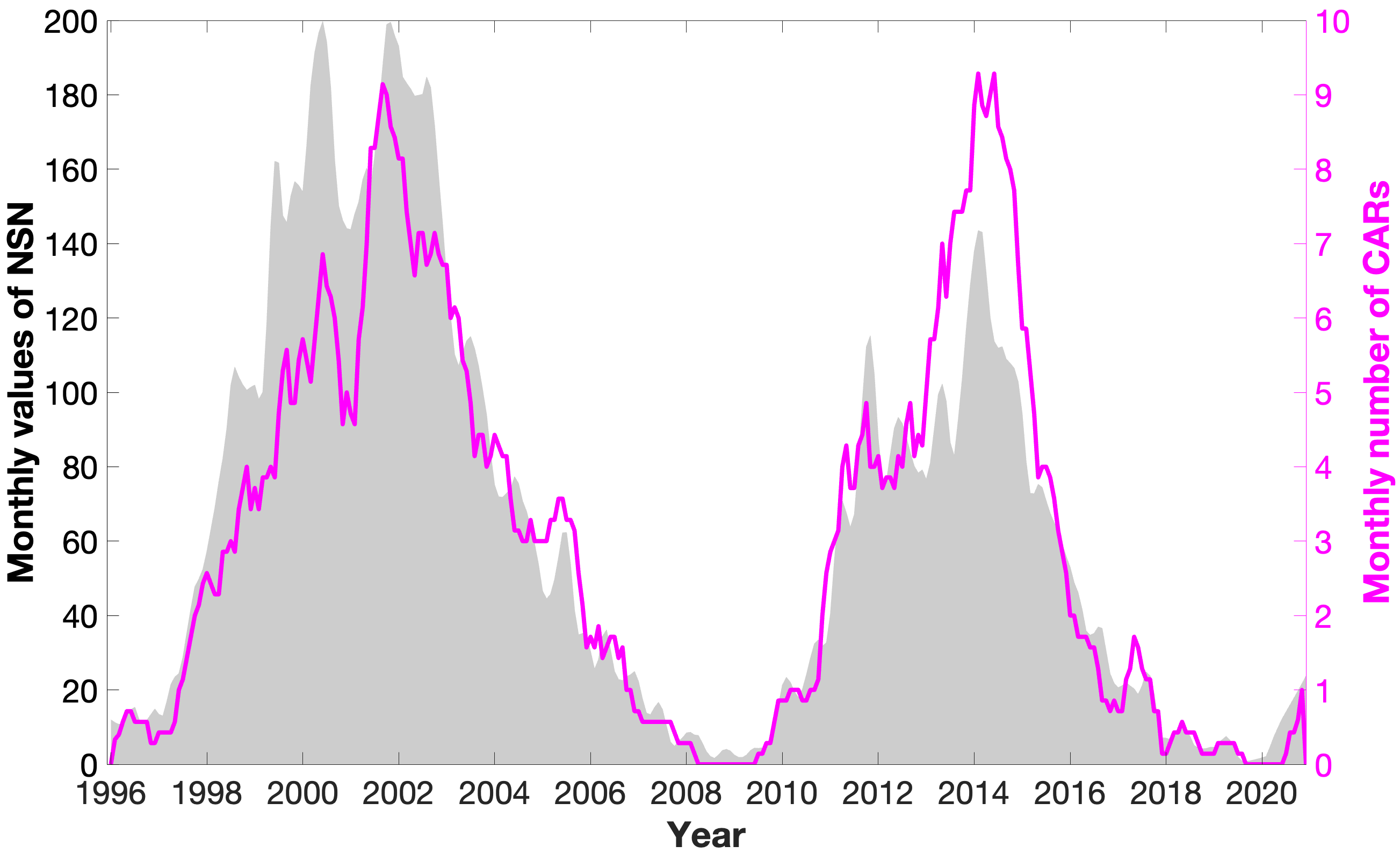}}
\caption{Variation in the monthly number of SARs (upper panel) and CARs (lower panel) in comparison with the monthly average values of the NSN (grey-shaded areas in both panels).}
\label{monthly}
\end{figure}

We studied the variation in the monthly number of SARs and CARs and compared the results to NSN. The monthly number of SARs (blue solid line in the upper panel) and CARs (purple solid line in the lower panel) are presented in Fig \ref{monthly}. The grey-shaded areas show the monthly average values of NSN in both upper and lower panels. All data in Fig \ref{monthly} have been smoothed using a seven-month moving average.

As can be seen from Fig \ref{monthly}, the monthly number of SARs and CARs closely follows the monthly values of NSN. The Pearson correlation coefficients between  SARs and NSN and between CARs and NSN are r = 0.97 and r = 0.90, respectively. In addition, both SARs and CARs show a double peak pattern during the solar maximum phases in SC 23 and 24; although SARs reached a maximum value during the first peak of the NSN, whereas CARs attained their maximum later and during the second peak of NSN in both SC 23 and 24.

\begin{figure}[h]
\resizebox{\hsize}{!}{\includegraphics{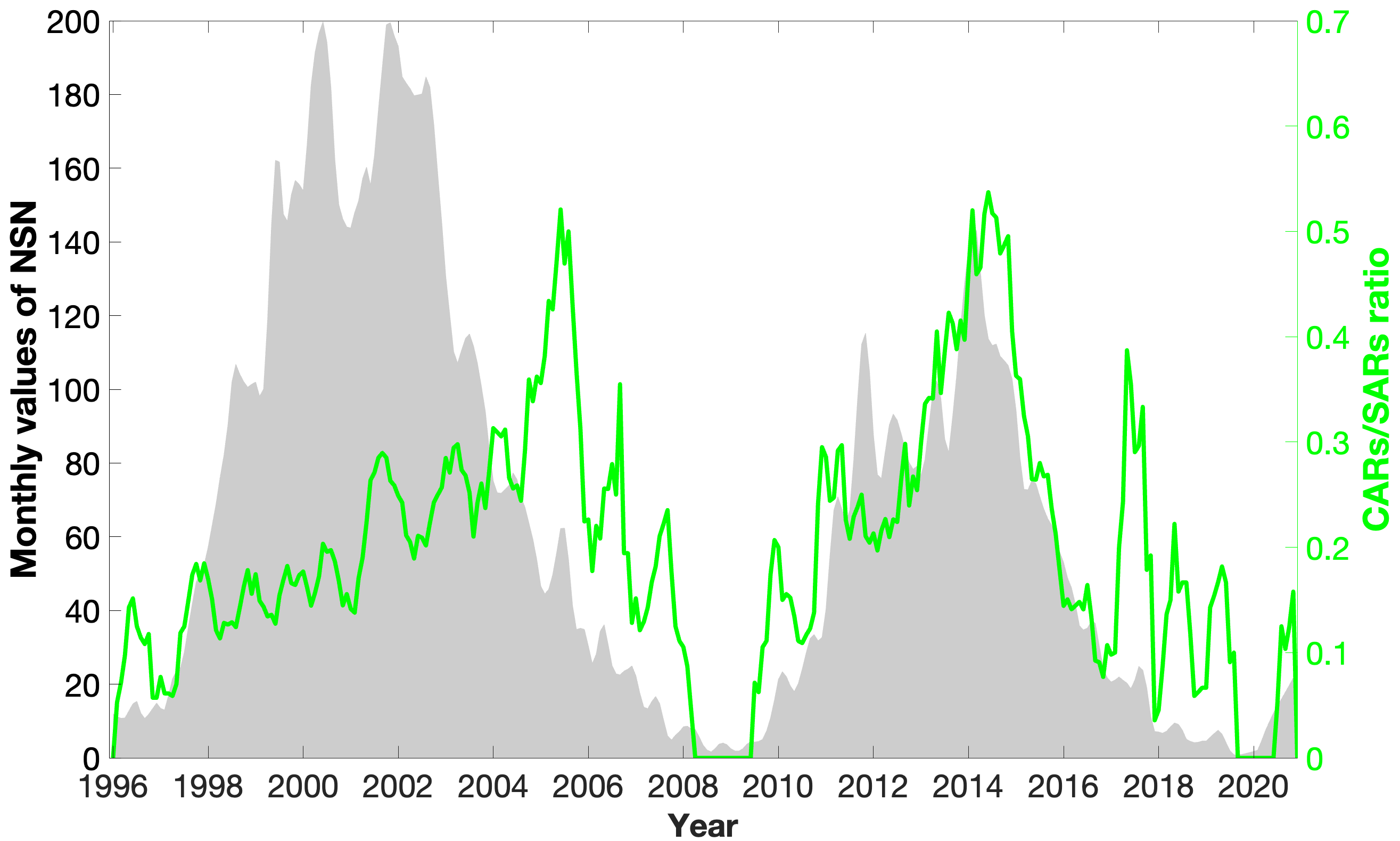}}
\caption{The monthly Ratio of the  number of CARs to SARs (green solid line) in comparison with the monthly values of NSN (grey-shaded area).}
\label{ratio}
\end{figure}

A comparison of the two panels of Fig \ref{monthly} indicates that the maximum monthly number of SARs dramatically decreases $46.3 \%$ from SC 23 to SC 24, while the maximum number of CARs is almost the same in both cycles, with only $1.5 \%$ increase from SC 23 to 24.

\begin{figure}[h]
\resizebox{\hsize}{!}{\includegraphics{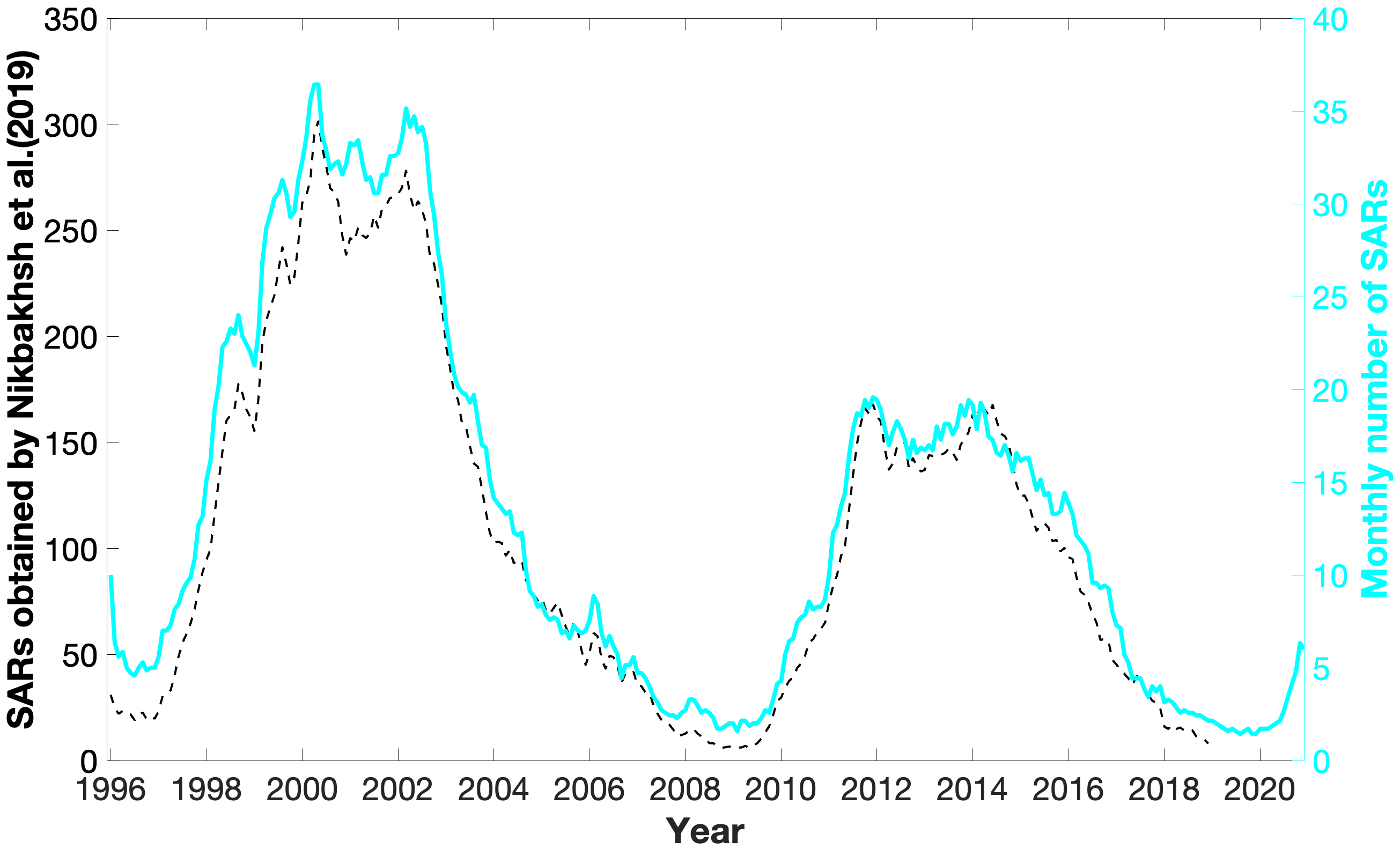}}
\\[1\baselineskip]
\resizebox{\hsize}{!}{\includegraphics{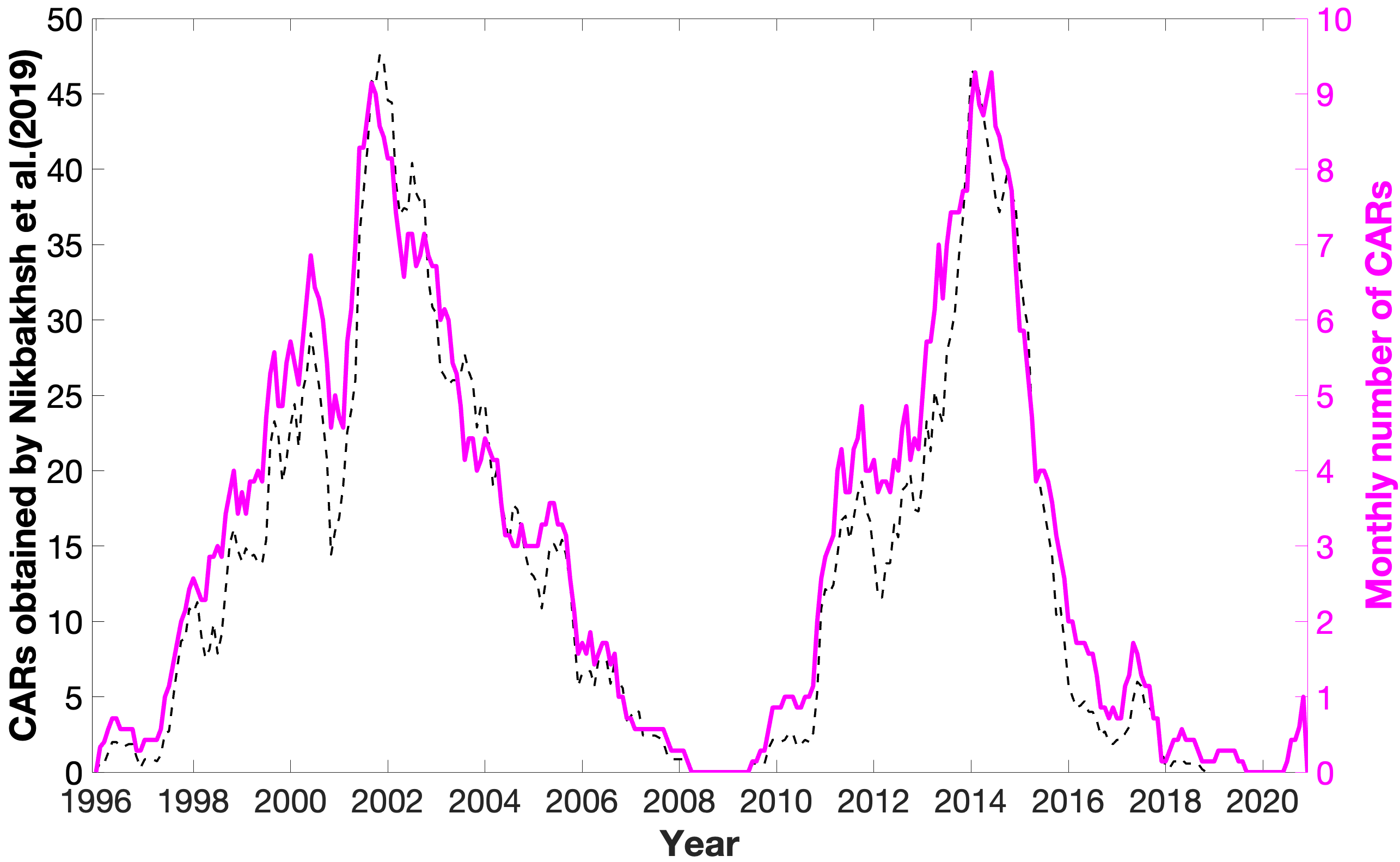}}
\caption{A Comparison between the monthly values of SARS (upper panel) and CARs (lower panel) to the results obtaind by \citet{Nikbakhsh2019} (presented by dashed lines in both panels)}
\label{comparision}
\end{figure}

Next, we computed the monthly ratio of the  number of CARs to SARs, which is presented in Fig \ref{ratio} (solid green line) in comparison to the monthly values of NSN (grey-shaded area). It can be seen from the figure that this ratio is ranging between 0 during solar minimum up to 0.6 during solar maximum in both cycles. In addition, it is evident that this ratio reaches its maximum during the declining phase of SC 23 and as well as in the late maximum phase of SC 24.
 
We also compared our results from the analysis of the monthly number of SARs and CARs (Fig \ref{monthly}) to the one obtained by \citet{Nikbakhsh2019}. Figure \ref{comparision} presents the results from this comparison. The results from this study are presented by solid lines while the results from \citet{Nikbakhsh2019} are displayed with dashed lines in both panels. By comparing two panels, one can see that, regardless of the method, ARs with more complex magnetic structures achieve their maximum value during the second peak of a cycle, whereas regions with simple structures attain a maximum value during the first peak.

We observed that the peaks in the monthly number of CARs were very similar in both cycle 23 and 24. Hence, we decided to conduct a similar analysis as \citet{Nikbakhsh2019} to closely investigate the monthly number of SARs and CARs during their peak activity. For that, we found out the maximum value of SARs and CARs in each cycle. The maximum of SARs was attained in April 2000 during SC 23 and in December 2011 during SC 24 while CARs peaked in September 2001 and February 2014 during cycle 23 and 24, respectively. Then we counted the monthly number of SARs and CARs for the period of two years before and after their maximum value in each cycle. Table \ref{maximum} presents the number of SARs and CARs during their peak activity periods in SC 23 and 24. As can be observed from the table, the monthly number of CARs in SC 23 was almost the same as the number in SC 24 with only $10.2  \%$ drop. On the other hand, the number of SARs dramatically decreased by $51.5 \%$ from SC 23 to SC 24.

\begin{table}[h]
\caption{Monthly number of SARs and CARs for the period of two years before and after their maximum values in SCs 23 and 24.}
\label{maximum}
\centering
\begin{tabular}{ c  c  c   c }
\hline\hline
 & Peak SC23 & Peak SC24 & Rate of change\\      
  & [Number] & [Number] & $[\%]$\\ 
 \hline 
 & & &\\ 
 SARs  & 1465  & 711 & 51.5  \\
  & & &\\ 
 & Peak SC23  & Peak SC24 & Rate of change \\ 
 & [Number] & [Number] & $[\%]$\\ 
 \hline 
  & & &\\ 
 CARs & 313 & 281 & 10.2 \\ 
  & & &\\ 
\hline 
\end{tabular}
\end{table}

\section{Discussion}
The magnetic complexity of an AR can evolve from a simple to a complex structure or the other way around during its lifetime. Hence, it is important to choose a suitable approach in order to study the magnetic complexity of ARs. We developed a new method called MEM and applied it on the AR data present in the SRS list (see section \ref{sec:DATA}). We discovered that  about $82 \%$ of all ARs were SARs for the study period. This result is in agreement with those obtained by \citet{Nikbakhsh2019} and \citet{Jaeggli2016} who reported that $88 \%$ and $84 \%$, respectively, of all ARs in their samples had simple structures. Moreover, we found out that $18.5 \%$ of all ARs were CARs. This result is also similar to the findings of \citet{Nikbakhsh2019} and \citet{Jaeggli2016}, which were $16 \%$ and $12 \%$, respectively. The very small difference between the numbers of identified complex regions is likely to be related to the different methods of classification.

We also studied the cyclic variation in number of SARs and CARs in SC 23 and 24. We demonstrated that the total number of SARs and CARs decreased by about $45.1 \%$ and $26 \%$, respectively, from SC 23 to 24. We  also found out that the ratio of CARs to SARs changed by $25.9 \%$ while the total number of ARs decreased by $41.9 \%$ from SC 23 to 24. 

By analysing our measured lifetime of ARs, we showed that SARs had a mean lifetime of 15.9 days in SC 23 and 15.1 in SC 24. On the other hand, CARs had a mean lifetime of 24.8 days in SC 23 and 22.3 days in SC 24. These results demonstrated that, on average CARs have a longer lifetime than SARs (about 8 days longer). In addition, we discovered that the mean lifetime of SARs and CARs had almost no variation from SC 23 to 24, despite the fact that these two cycles are very different; hence, we suggest that the lifetime of SARs and CARs is independent of the solar cycle. 

A detailed analysis on the  lifetime data revealed that SARs had bimodal distribution around 12 and 26 days (Fig\ref{lifetime}). It seems that this bimodal distribution is due to selection effect which is the combination of the Sun's rotation period and visibility. This bias will, of course, influence the estimates of the mean life times. In addition, we demonstrated that CARs had a left-skewed distribution with $87.8 \%$ of the regions having  a lifetime longer than 20 days (Fig \ref{lifetime}). 

Further analysis on the shape of the distributions for both SARs and CARs revealed that their distribution shapes were similar in both cycles. Hence, we suggest that the distribution shape of lifetime of SARs and CARs is independent of the solar cycle.

A detailed analysis on the magnetic evolution of CARs showed that $93.8\%$ of these regions appeared to the photosphere while possessing a simple structure ($\alpha$ or $\beta$) before they transform into complex structures (e.g., $\beta\gamma$ or $\beta\gamma\delta$). In this respect, we demonstrated that the Growth phase for these regions had a mean value of 3.2 days which is about 1/7 of their lifetime. Moreover, we found out that the mean value of the Main phase was 4.8 days for these regions. Hence, we can conclude that on average CARs possess a complex structure for about one-fifth of their lifetime. This finding can be used to forecast the occurrence of solar flares and profoundly improve space weather prediction.

We also calculated the monthly ratio of the number of CARs to SARs. We demonstrated that this ratio varied between 0 to  0.6 with a mean value of 0.2 (Fig \ref{ratio}). The ratio reached a maximum value during the declining phase of SC 23 and as well as in the later maximum phase of SC 24; 
This finding agrees with \citep{Nikbakhsh2019}, who hypothesized that the ARs are formed by a process which is a result of a competition between the large-scaled dynamo (LSD) and the small-scaled dynamo (SSD). They suggested that during solar maximum, LSD is dominant, and this gives a preference to the formation of SARs. On the other hand, the relative role of SSD increases during the declining phase of a cycle, which give a preference to the formation of CARs.

We investigated the monthly numbers of SARs and CARs during their peak activity periods in both SC 23 and 24. On this point, we showed that the rate of change for the total number of CARs was only $10.2 \%$; on the contrary, this rate was  $51.5 \%$ for SARs (Table \ref{maximum}). A similar analysis was conducted by \citet{Nikbakhsh2019}, who discovered that the rate of change for complex and simple magnetic structures were $15 \%$ and $51 \%$, respectively, from SC 23 to SC 24.  \citet{Nikbakhsh2019} suggested that SSD is the responsible mechanism for transforming ARs to more complex regions, and this mechanism seems to be cycle independent. Therefore, we observe a smaller rate of change in the number of CARs. 
 
\section{Conclusions}
We examined the magnetic complexity of ARs using a newly developed method that enabled us to investigate the full magnetic evolution of both simple and complex structures over the period January 1996 to December 2020. We analyzed 4841 individual ARs, of which 898 ($19 \%$) were classified as CARs. Our results show that, on average, CARs have a lifetime of approximately 24 days, and that the majority of them ($94 \%$) initially appear with a simple magnetic structure. On average, these regions transition into complex magnetic structures approximately three days after their initial appearance on the photosphere, a period we define as the growth phase.

The key discovery of this study is that ARs typically develop complex magnetic structures approximately three days after their initial emergence on the solar photosphere, a period we define as the growth phase. This phase likely reflects the build-up of magnetic complexity associated with increased eruptive potential. Understanding the growth phase is essential for identifying the physical mechanisms that drive the transition from simple to complex magnetic morphology.

Following the growth phase, the magnetic morphology of ARs typically becomes complex. This state persists for approximately five days—about one-fifth of the region’s total lifetime, and is defined as the main phase. The main phase is exclusive to CARs and distinguishes them from SARs. We hypothesize that the majority of solar eruptions, such as flares and CMEs, occur during this phase when magnetic complexity peaks. Further statistical studies are needed to quantify the relationship between AR magnetic morphology and eruptive activity. Such information is essential for developing more accurate models to forecast space weather events.

\section*{Acknowledgments}
We acknowledge the financial support by the Academy of Finland of the SOLSTICE (project 324161), Earth-Space Research Ecosystem E2S (project nro 336719 and 337663) and ReSoLVE Center of Excellence (project nro 272157 and 307411). We would like to thank the National Oceanic and Atmospheric Administration (NOAA) and the Heliophysics Integrated Observatory (HELIO) for providing the solar active region and sunspot data.

\bibliographystyle{aa}
\bibliography{Nikbakhsh}

\begin{thebibliography}{21}
\expandafter\ifx\csname natexlab\endcsname\relax\def\natexlab#1{#1}\fi

\bibitem[{{Andrus}(2013)}]{Andrus2013}
{Andrus}, D. 2013, {Air Force Weather Agency Manual},
  \url{https://ngdc.noaa.gov/stp/space-weather/online-publications/miscellaneous/afrl_publications/afwaman15-1_space-environmental-observations.pdf},
  page 37, Air Force Departmental Publishing Office

\bibitem[{{Chen} {et~al.}(2011){Chen}, {Wang}, {Shen}, {Ye}, {Zhang}, \&
  {Wang}}]{Chen2011}
{Chen}, C., {Wang}, Y., {Shen}, C., {et~al.} 2011, Journal of Geophysical
  Research (Space Physics), 116, A12108

\bibitem[{{Clette} {et~al.}(2016){Clette}, {Lef{\`e}vre}, {Cagnotti},
  {Cortesi}, \& {Bulling}}]{Clette2016}
{Clette}, F., {Lef{\`e}vre}, L., {Cagnotti}, M., {Cortesi}, S., \& {Bulling},
  A. 2016, \solphys, 291, 2733

\bibitem[{{Cortie}(1901)}]{Cortie1901}
{Cortie}, A.~L. 1901, \apj, 13, 260

\bibitem[{{Guo} {et~al.}(2014){Guo}, {Lin}, \& {Deng}}]{Guo2014}
{Guo}, J., {Lin}, J., \& {Deng}, Y. 2014, \mnras, 441, 2208

\bibitem[{{Hale}(1908)}]{Hale1908}
{Hale}, G.~E. 1908, \apj, 28, 315

\bibitem[{{Hale} {et~al.}(1919){Hale}, {Ellerman}, {Nicholson}, \&
  {Joy}}]{Hale1919}
{Hale}, G.~E., {Ellerman}, F., {Nicholson}, S.~B., \& {Joy}, A.~H. 1919, \apj,
  49, 153

\bibitem[{{Hathaway}(2015)}]{Hathaway2015}
{Hathaway}, D.~H. 2015, Living Reviews in Solar Physics, 12, 4

\bibitem[{{Howard}(1989)}]{Howard1989}
{Howard}, R.~F. 1989, Solar physics, 123, 271

\bibitem[{{Howard}(1991)}]{Howard1991}
{Howard}, R.~F. 1991, Solar physics, 131, 239

\bibitem[{{Ireland} {et~al.}(2008){Ireland}, {Young}, {McAteer}, {Whelan},
  {Hewett}, \& {Gallagher}}]{Ireland2008}
{Ireland}, J., {Young}, C.~A., {McAteer}, R.~T.~J., {et~al.} 2008, \solphys,
  252, 121

\bibitem[{{Jaeggli} \& {Norton}(2016)}]{Jaeggli2016}
{Jaeggli}, S.~A. \& {Norton}, A.~A. 2016, \apjl, 820, L11

\bibitem[{{McIntosh}(1990)}]{McIntosh1990}
{McIntosh}, P.~S. 1990, \solphys, 125, 251

\bibitem[{{Nagovitsyn} {et~al.}(2019){Nagovitsyn}, {Ivanov}, \&
  {Skorbezh}}]{Nagovitsyn2019}
{Nagovitsyn}, Y.~A., {Ivanov}, V.~G., \& {Skorbezh}, N.~N. 2019, Astronomy
  Letters, 45, 396

\bibitem[{{Nikbakhsh} {et~al.}(2019){Nikbakhsh}, {Tanskanen}, {K{\"a}pyl{\"a}},
  \& {Hackman}}]{Nikbakhsh2019}
{Nikbakhsh}, S., {Tanskanen}, E.~I., {K{\"a}pyl{\"a}}, M.~J., \& {Hackman}, T.
  2019, \aap, 629, A45

\bibitem[{{Schrijver} \& {Zwaan}(2000)}]{Schrijver2000}
{Schrijver}, C.~J. \& {Zwaan}, C. 2000, {Solar and Stellar Magnetic Activity}

\bibitem[{{Stenning} {et~al.}(2013){Stenning}, {Lee}, {van Dyk}, {Kashyap},
  {Sandell}, \& {Young}}]{Stenning2013}
{Stenning}, D.~C., {Lee}, T.~C.~M., {van Dyk}, D.~A., {et~al.} 2013,
  Statistical Analysis and Data Mining: The ASA Data Science Journal, 6, 329

\bibitem[{{Subramanian} \& {Dere}(2001)}]{Subramanian2001}
{Subramanian}, P. \& {Dere}, K.~P. 2001, \apj, 561, 372

\bibitem[{{Tang} {et~al.}(1984){Tang}, {\textit{Howard}}, \&
  {Adkins}}]{Tang1984}
{Tang}, F., {\textit{Howard}}, R., \& {Adkins}, J.~M. 1984, \solphys, 91, 75

\bibitem[{{van Driel-Gesztelyi} \& {Green}(2015)}]{Van2015}
{van Driel-Gesztelyi}, L. \& {Green}, L.~M. 2015, Living Reviews in Solar
  Physics, 12, 1

\bibitem[{{Waldmeier}(1938)}]{Waldmeier1938}
{Waldmeier}, M. 1938, \zap, 16, 276

\end{thebibliography}
\end{document}